\documentstyle[12pt,dpf]{article}

\newcommand{\cM}{{\cal M}}

\title{\bf QCD RESULTS FOR NUCLEON COMPTON SCATTERING}
\author{A.S. KRONFELD$^a$ and B. NI\v{Z}I\'{C}$^b$ \\[0.3cm]
{\it $^a$Theoretical Physics Group, Fermi National Accelerator
Laboratory,} \\
{\it P.O. Box 500, Batavia, IL 60510, USA} \\[0.5em]
{\it $^b$Rudjer Bo\v{s}kovi\'{c} Institute,} \\
{\it P.O. Box 1016, HR-41001 Zagreb, Croatia}}
\abstract{We present QCD results for the exclusive processes
$\gamma{\rm N}\rightarrow\gamma{\rm N}$ ($\rm N=p,\;n$) at large
momentum transfer and compare them to data for the proton.}
\begin{document}
\vfill 
\maketitle
\vfill \newpage 
The amplitude for a wide-angle exclusive process is given by the
convolution of distribution amplitudes $\phi$ summarizing soft,
hadronic physics and a hard-scattering amplitude $T$
of collinear, constituent partons.\cite{Lep80}
For nucleon Compton scattering
\begin{equation}\label{h-amplitude}
\cM_{hh'}^{\lambda\lambda'}(s, t) =
\sum_{d,i}\int[dx][dy]\,
\phi_i(x_1,x_2,x_3) T_i^{(d)}(x,h,\lambda;y,h',\lambda')
\phi^*_i(y_1,y_2,y_3),
\end{equation}
where $x_a$ ($y_a$) are momentum fractions of the quarks in the initial
(final) state proton, the $\lambda$'s and $h$'s are helicities, and $i$
and $d$ label proton Fock states and Feynman diagrams.
Both $\phi$ and $T$ depend on a factorization scale $\mu$, but since we
work to lowest order, we shall neglect this dependence.
The integral in eq.~(\ref{h-amplitude}) is somewhat like a loop integral.
In particular, internal partons can go on mass shell for
certain $(x,y)$, producing an imaginary part. 
The amplitude is still infrared safe,\cite{Far89} because on-shell
internal partons move in a direction that tears the nucleon apart.

This paper summarizes our results for the cross sections for nucleon
Compton scattering,\cite{Kro91} which is the simplest experimentally
accessible process with an imaginary part.
The predictions for polarized cross sections and phase of the amplitude
can be verified in $e{\rm p}$ collisions,\cite{Far90} because Compton
scattering with a virtual incident photon also contributes to the
reaction $e{\rm N}\rightarrow e{\rm N}\gamma$.
This would be interesting, because the non-zero phase is a non-trivial
prediction of perturbative QCD.
For unpolarized proton Compton scattering there is wide-angle
data\cite{Shu79} with center-of-mass energy-squared
$4.6\,{\rm GeV}^2<s<12.1\,{\rm GeV}^2$.
Since the distribution amplitude is not known,
we will present results using four distribution amplitudes suggested by
QCD sum rules.\cite{Che84,Che89,Kin87,Gar86}
These $\phi$'s implicitly assume that all moments
except the first six vanish.
The validity of such an assumption at accessible values of
$s$ remains to be tested.

The space allotted permits no discussion of the calculation.
The difficult aspects of the calculation are technical, especially
coping with the on-shell singularities in the momentum-fraction
integrals.
Wherever possible, we integrate singular integrands analytically.
For some diagrams poles remain in the domain of numerical integration,
and we use the technique developed in \refcite{Niz87}.
For details, please consult \refcite{Kro91}.

Polarized cross sections and phases are presented in \refcite{Kro91}.
Here we present only unpolarized cross sections.
Our results for $s^6d\sigma/dt$ are plotted in Fig.~\ref{plots}
for the proton and neutron.
Four different distribution amplitudes are shown,
CZ\cite{Che84} (dashed lines), COZ\cite{Che89} (solid lines),
KS\cite{Kin87} (dotted lines), and GS\cite{Gar86} (dot-dashed lines).
Fig.~\ref{plots}(a) also includes the experimental data.\cite{Shu79}
The agreement is encouraging, especially in light of the uncertainties
discussed below.
\begin{figure}
\vspace{9.5cm}
\caption[plots]{%
Differential cross sections for (a) protons and (b) neutrons.
The experimental data\cite{Shu79} in (a) are at
$s=4.63$ GeV (circles), $s=6.51$ GeV (triangles),
$s=8.38$ GeV (squares), $s=10.26$ GeV (five-pointed stars),
and $s=12.16$ GeV (asterisk).}\label{plots}
\end{figure}

According to the dimensional counting rules,\cite{Bro73,Lep80}
$s^6d\sigma/dt$ should be independent of $s$.
In QCD several effects lead to deviations from this rule.
First, there is the running of the QCD coupling constant, which we have
fixed at $\alpha_{\rm S}=0.3$, as in other
calculations.\cite{Mai88,Far90}
The cross section is sensitive to this choice, because it is
proportional to $\alpha_{\rm S}^4$.
Second, there is the running of the distribution amplitude.
The spread of the curves gives a qualitative estimate of this effect.
Third, there are mass effects; the nucleon masses is not negligible
compared to the photon energies in \refcite{Shu79}.
Finally, there are higher twist effects, coming from scattering of
non-valence Fock states.

The largest systematic uncertainty in our predictions comes from
the nucleon decay constant.
The cross section is proportional to $f_{\rm N}^4$, and we have used the
value $f_{\rm N}=(5.2\pm0.3)\times10^{-3}\;{\rm GeV}^2$ suggested by QCD
sum rules.\cite{Che84,Kin87}
Accepting the error estimate at face value yields a 23\% uncertainty in
the cross sections.
On the other hand, using the value suggested by quenched lattice
QCD,\cite{Mar89} $f_{\rm N}=(2.9\pm0.6)\times10^{-3}\;{\rm GeV}^2$,
would reduce the cross section by a factor of 9.

Since \refcite{Kro91} was finished, there have
been theoretical and experimental developments of interest.
Neglecting the transverse size of the hadron suggests the
factorization scale\cite{Lep80} $\mu=\min\{x_i\}Q$, which is
problematic near $x_a=0$.
Better estimates of the soft regime\cite{BSL89} show that the
transverse size introduces a natural infrared cutoff to factorization,
$\mu=\max\{q_g(x,y), 1/|b_\perp|\}$.
Calculations taking these ``Sudakov'' effects into account exist for
the proton form factor\cite{Li92} and
$\gamma\gamma\rightarrow{\rm p\bar{p}}$ (\refcite{Hye92}).
A proposed experiment at SLAC\cite{Hyd92} promises to acquire high
statistics at high energy for $\gamma{\rm p}\rightarrow\gamma{\rm p}$
and $\gamma{\rm p}\rightarrow\pi{\rm N}$.
It makes sense, therefore, to improve our results along the lines of
\refcite{Li92,Hye92} and to extend the calculations to exclusive
pion photoproduction.

\vspace{1.0em} 
We thank G.P. Lepage for useful discussions.

\end{document}